\documentclass[aps,prl,amsmath,amssymb,onecolumn,nofootinbib,showpacs,tightenlines,12pt]{revtex4}

\usepackage{epsfig,graphicx}
\usepackage{bm}
\newcommand{\be}{\begin{equation}}
\newcommand{\ee}{\end{equation}}
\newcommand{\bea}{\begin{eqnarray}}
\newcommand{\eea}{\end{eqnarray}}

\begin{document}
\title{Cylindrically Symmetric Scalar Field  and it's Lyapunov stability in General Relativity }
\author{H.R.Rezazadeh}
\email{h-rezazadeh@kiau.ac.ir}
 \affiliation{Department of Mathematics,Faculty of Science, Islamic Azad
University,Karaj branch,Karaj, P.O. Box 31485-313, Karaj, Iran}

\begin{abstract}

In this paper we found an  Exact solution for massless scalar field
with cosmological constant.This exact solution generalized the
Levi-Civita vacuum solution\cite{8} to a massless scalar field,with
a cosmological constant term.This solution in the absence of the
Cosmological constant recovers the spacetime of a massless scalar
field with cylindrical symmetry(\emph{Buchdahl metric}\cite{2}).Also
if the scalar field disappears, the spacetime is a representation of
de-Sitter space.We prove that the form of the metric's function
which was purposed in \cite{1} is valid even if we assume a general
form.Too we show that in which conditions this solution satisfies
energy conditions. Finally the validity of focusing theorem is
proved.
\end{abstract} \maketitle
\section{\label{sec:level1}Introduction}

In the \cite{1} the authors acclaimed that they were succeed in
solving the problem of finding an exact solution of Einstein field
equations for a massless scalar field with cylindrical symmetry in
the presence of a cosmological constant\cite{1}.They obtained a new
two parameter exact solution (LB meric)\footnote{ This is an
abbreviation to the memories of H. A. Buchdahl and the presence of
the cosmological constant term} which recovers at least one member
of Buchdahl family \cite{2} in the absence of Cosmological constant
and also $LC\Lambda$ family\cite{3}.Theire  work in [1] completed
previous works as [3,6,8,11].In that work the authors solved the
field equations in a special case and no thing was stated about the
general exact solution. Now we applied a method for solving field
equations and show that we can construct a general two parameters
exact solution which in the absence of  the cosmological constant
recover Buchdahl solution. We show that this system of differential
equations posses only those solutions that the authors stated in
\cite{1}.Later a class of solutions of Einstein field equations is
investigated for a cylindrically symmetric spacetime when the source
of gravitation is a perfect fluid [4].As a historical note we added
here that, our solution with Cosmological constant and scalar field
is a generalization of Levi-Civita family[7,8].Previously the exact
solution of Einstein field equations with a cosmological constant
term were found by Linet [10]and in a more efficient form by Tian
[11].The singularity problem in a family of cylindrically symmetric
spacetimes which is described by the collapsing of scalar fields was
discussed by Wang and Frankel [9].Senovilla presented an explicit
exact solution of Einstein's equations for an inhomogeneous dust
universe with cylindrical symmetry[5].In this work first we prove
that the general solution for field equations only is one, which was
stated in \cite{1}.Then we investigated the validity of Energy
conditions and focusing theorem for our solution.Also a short
section is present which discusses the stability of exact solution
under small perturbations via via Lyapunov exponents
method[27,28,29].

\section{i:Field equations}
 We begin with a general cylindrically symmetric metric
in Weyl coordinates   $(t,r,\varphi,z)$,
\begin{eqnarray}
ds^{2}=-e^{u(r)}dt^2+dr^2+e^{v(r)}d\varphi^2+e^{w(r)}dz^2
\end{eqnarray}
Field equation for a massless minimally coupled scalar field in the
presence of a cosmological constant term $\Lambda$ is reading as
\footnote{I will mostly use natural units $\hbar = c = 1$  and $8\pi
G = 1$.}:
\begin{eqnarray}
R_{\mu\nu}-\Lambda g_{\mu\nu}=\phi_{;\mu}\phi_{;\nu}
\end{eqnarray}
 We labeled metric functions as $u_{i}=\{u(r),v(r),w(r)\},\phi\equiv \phi(r)$ , $\acute{h}=\frac{d h}{d
 r}$.
 In terms of these functions we can rewrite field equation (2) in the
 following succinct forms:
\begin{eqnarray}
2u_{i}''+u_{i}'\sum^{3}_{j=1}u_{j}'-4 \Lambda=0, i=\{1,2,3\}\\
2\sum^{3}_{j=1}u_{j}''+\sum^{3}_{j=1}u_{j}'^{2}-4 \Lambda=4\phi'^{2}
\end{eqnarray}
Now we write equation (3) in the following simple form:
\begin{eqnarray}
\frac{d}{dr}(u_{i}'e^{f})=2 \Lambda e^{f}, i=\{1,2,3\}
\end{eqnarray}
Where in it $f=\frac{\sum^{3}_{j=1}u_{j}}{2}$. If we do summation on
$i=1,2,3$  in (5) we can write a differential equation for new
function $f=f(r)$,
\begin{eqnarray}
\frac{d}{dr}(f'e^{f})=3 \Lambda e^{f}
\end{eqnarray}
In terms of this new function the equation (4) converted to the
following equation:
\begin{eqnarray}
3f''+f'^2-7\Lambda=\phi'^{2}
\end{eqnarray}
Substituting $f''$ from (6) in (7) we obtain the following integral
for $\phi$:
\begin{eqnarray}
\phi=\pm\sqrt{2}\int dr\sqrt{\Lambda-f'^2}
\end{eqnarray}
Thus if we solve the equation (6) we can obtain both metric
functions $u_{i}$ and $\phi$.
\subsection{About the stability of the field equations}
 Now we investigate the stability of the field
equations (3,4) under small pertubations.First we note that instead
of working with the system (3,4) we can treat the (6,8) as the field
equations.in linear approximation (6) becomes
\begin{eqnarray}\nonumber
f''\approx3 \Lambda
\end{eqnarray}
The general solution for this simple ODE is
\begin{eqnarray}\nonumber
f(r)=\frac{3 \Lambda}{2}r^2+c_{1}
\end{eqnarray}
If we take the de-Sitter radius as $a=\sqrt{\frac{3}{\Lambda}}$ this
function is nothing but the usual metric function of the Asymptotic
de-Sitter.If we want to check the eigenvalues of the linearized
matrix for system of field equations we must change the system (3,4)
to a higher rank first order system.For this, we introduce the
following set of new variables,
\begin{eqnarray}\nonumber
\acute{u_{i}}=x_{i} i=\{1,2,3\},\acute{\phi}=y
\end{eqnarray}
Then we have
\begin{eqnarray}\nonumber
2x_{i}'+x_{i}\sum^{3}_{j=1}x_{j}-4 \Lambda=0, i=\{1,2,3\}\\\nonumber
2\sum^{3}_{j=1}x_{j}'+\sum^{3}_{j=1}x_{j}^{2}-4 \Lambda=4y'^{2}
\end{eqnarray}
For a stationary point in the phase space we must set all first
order derivatives equal to zero,
\begin{eqnarray}\nonumber
x_{i}\sum^{3}_{j=1}x_{j}-4 \Lambda=0, i=\{1,2,3\}\\\nonumber
\sum^{3}_{j=1}x_{j}^{2}-4 \Lambda=0,y=0
\end{eqnarray}
In the first equation if we summing on the indices and comparing
both of them we obtain
\begin{eqnarray}\nonumber
x_{i}=2/a, i=\{1,2,3\},y=0
\end{eqnarray}
We perturb the field equations as
\begin{eqnarray}\nonumber
x_{i}=2/a+\delta x_{i} i=\{1,2,3\},y=\delta y
\end{eqnarray}
Expanding the field equation in these perturbations up to first
order and taking to the mind to having an asymptotic stability the
real part of the eigenvalues must be negative.In the case of our
field equations the matrix posses negative purely real
eigenvalues,thus according to the Lyapunov theorem this system is
stable.
\section{ii:Exact solutions}
In this section we investigate all possible solutions for (6),(8).

 The exact solution for ordinary differential equation (6) is\footnote{$\log(x)=\int^{x}_{1}\frac{d
\zeta}{\zeta}$}:
\begin{eqnarray}
f(r)=-\sqrt{3\Lambda}r+\frac{1}{2}\log(\frac{1}{12\Lambda}(c_{1}e^{2\sqrt{3\Lambda}r}-c_{2})^2)
\end{eqnarray}
We keep the relation (9) as a constraint for all solutions of
(6).Substituing this function in (5) we can write the following
general form of metric functions:
\begin{eqnarray}
u_{i}=-\frac{\alpha_{i}}{3}\sqrt{\frac{3}{\Lambda}}\frac{1}{\sqrt{c_{1}c_{2}}}\tanh^{-1}(\sqrt{\frac{c_{1}}{c_{2}}}
e^{\sqrt{3\Lambda}r})+\frac{2}{3}\log(\frac{c_{1}e^{2\sqrt{3\Lambda}r}-c_{2}}{e^{\sqrt{3\Lambda}r}})+\beta_{i},i=1,2,3
\end{eqnarray}
We note here that if the solutions (10) satisfy in (9) , then we
have $\sum^{3}_{j=1}\alpha_{j}=0$.Thus the final form of metric
functions in :
\begin{eqnarray}
u_{i}=\frac{2}{3}\log(\frac{c_{1}e^{2\sqrt{3\Lambda}r}-c_{2}}{e^{\sqrt{3\Lambda}r}})+\beta_{i},i=1,2,3
\end{eqnarray}
Later we  will determine the coefficients $c_{1},c_{2}$. The
condition $\sum^{3}_{j=1}\beta_{j}=-\frac{1}{2}\log(12\Lambda)$ may
be satisfied by adding a constant $\beta_{j}$ to each metric
function which can be absorb in the non radial coordinates.In the
languages of potential theory we can choose a new gauge for
potential functions $u_{i}(r)$ .The reason is that in cylindrical
spacetimes we know that the  solving of the Einstein field equations
can be reduced to solving a system of potential equations\cite{7}. A
better option is the  comparison of (11) with the previous metric
function in \cite{1}:
\begin{eqnarray}
u_{i}=u(r)=\pm\sqrt{\frac{\Lambda}{3}}r+ \frac{2}{3}\log(1+\xi^2
e^{2\sqrt{3\Lambda}r}),i=1,2,3
\end{eqnarray}
As it was attend that in \cite{1}by the authors , only the minus
sign in metric function is consistent with scalar field equation of
motion $\Box \phi=0$ .In terms of a new parameter
$\xi=-i\sqrt{\frac{c_{1}}{c_{2}}}\in \mathbb{R}$- which was defined
in \cite{1}- we must choose the constant $\beta_{j}$ as the
following:
\begin{eqnarray}
\beta_{j}=-\frac{2}{3}\log(-c_{2}),i=1,2,3
\end{eqnarray}

Obviously if $c_{1}\in \mathbb{R}$, then $c_{2}\in \mathbb{C}$
(purely imaginary) and consequently $\beta_{j}$ is real.
\section{iii:Validity of energy condition in  LB metric}
The origin of the null energy condition \emph{(\emph{NEC)}} and of
the strong energy conditions \emph{(SEC)} is the Raychaudhuri
equation together with the requirement that the gravity is
attractive for a spacetime manifold endowed with a metric
$g_{\hat{\mu}\hat{\nu}}$ \footnote{This is a Tetrad's representation
of the metric}[24].

The classical energy conditions of general relativity, to the extent
that one believes that they are a useful guide [12, 13], allow one
to deduce physical constraints on the behavior of matter fields in
strong gravitational fields or cosmological geometries. These
conditions can most easily be stated in terms of the components of
the stress energy tensor $T^{\hat{\mu}\hat{\nu}}$  in an orthonormal
frame. Ultimately, however, constraints on the stress-energy are
converted, via the Einstein equations, to constraints on the
spacetime geometry. \footnote{ In particular in a FRW spacetime one
is ultimately imposing conditions on the scale factor and its time
derivatives}For a perfect fluid cosmology, and in terms of pressure
and density, the so-called \emph{Null, Weak, Strong and Dominant
energy} conditions
reduce to [14]:\\\\
 \emph{NEC:}$p_{i}+\rho\geq 0$\\
\emph{ WEC:}This specializes to the \emph{NEC} plus $\rho\geq 0$\\
\emph{SEC:} This specializes to the NEC plus $\sum p_{i}+\rho\geq 0$\\
\emph{DEC:} $\rho\geqslant |p_{i}|$.\\\\
 Note particularly that in FRW models of Universe, the condition
 \emph{SEC} is independent of the space extrinsic curvature k. Now, \emph{DEC} implies \emph{WEC}
implies\emph{ NEC}, and SEC implies \emph{NEC}, but otherwise the
\emph{NEC, WEC}, \emph{SEC}, and \emph{DEC} are mathematically
independent assumptions. In particular, the SEC does not imply the
\emph{WEC}. Violating the \emph{NEC} implies violating the
\emph{DEC, SEC,} and \emph{WEC} as well [14].
 Note that ideal
relativistic fluids satisfy the \emph{DEC}, and certainly all the
known forms of normal matter encountered in our solar system satisfy
the \emph{DEC}. With sufficiently strong self-intereactions
relativistic fluids can be made to violate the \emph{SEC} (and
\emph{DEC}); but classical relativistic fluids always seem to
satisfy the \emph{NEC}. Most classical fields (apart from
non-minimally coupled scalars) satisfy the \emph{NEC}. Violating the
\emph{NEC} seems to require either quantum physics (which is    `
unlikely to be a major contributor to the overall cosmological
evolution of the universe) or non-minimally coupled scalar fields
(implying that one is effectively adopting some form of
scalar-tensor gravity). Using this dynamical formulation of the
energy conditions, Santos et al. [15] derive some bounds, for the
special case $k = 0$(flat cosmology), on the luminosity distance
$d_{L}$ of supernovae, and then contrast this with the legacy [16,
17] and gold [18] datasets. In reference [19] bounds on the distance
modulus are presented for general values of $k$  while in reference
[20] they
concentrate on the lookback time.\\
 Due to the lack of satisfactory dark energy models, many model-independent methods were proposed to
study the properties of dark energy and the geometry of the
universe[21,22,23]. Another very interesting and model-independent
approach is to consider the energy conditions [24,25]. Recently
Energy Conditions and Stability in $f(R)$ theories of gravity with
non-minimal coupling to matter  is discussed  and determined the
bounds from the energy conditions on a general $f(R)$ functional
form in the framework of metric variational approach.[26].

\section{iv: Checking the focusing theorem }
 In this section first we
state the focusing theorem in the manner of congruences both in time
like and null cases. Then we checked that our metric which is
constructed from a messless cylindrically symmetric scalar field in
the presence of a cosmological constant satisfies this theorem or
not?.
\subsection*{a:Focusing theorem in General Relativity}
Let a congruence of time like geodesics be hypersurface orthogonal,
means there is exist a vector field $u^{\alpha}$ (time like, space
like or null and not necessarily geodesic).We called it a
hypersurface orthogonal if $\omega_{\alpha\beta}\equiv
u_{[\alpha;\beta]}=0$.It means that there exist a scalar field
$\Phi$ such that $u_{\alpha}\varpropto \Phi_{,\alpha}$ and let the
(SEC) hold. So that from Einstein field equations
\begin{eqnarray}\nonumber
R_{\alpha\beta}u^{\alpha}u^{\beta}\geqslant 0.
\end{eqnarray}
From Raychaudhuri equation \cite{8} implies:
\begin{eqnarray}
\frac{d\theta}{d\tau}=-\frac{1}{3}\theta^2-\sigma^{\alpha\beta}\sigma_{\alpha\beta}-R_{\alpha\beta}u^{\alpha}u^{\beta}\leqslant
0
\end{eqnarray}
Where in it, $\theta=u^{\alpha}_{;\alpha}=B^{\alpha}_{\alpha}$ is
the expansion scalar ,
$\sigma_{\alpha\beta}=B_{(\alpha\beta)}-\frac{1}{3}\theta
h_{\alpha\beta}$ the shear tensor, $h_{\alpha\beta}$ the transverse
part of $g_{\alpha\beta}$ \cite{9} (which is purely 'spatial'). The
expansion must therefor decrease during the congruence's evaluation.
Focusing theorem stated that  \emph{an initially diverging
($\theta>0$) congruence will diverge less rapidly in the future,
while an initially converging ($\theta<0$) congruence will converge
more rapidly in the future}. The physical interpretation is that
gravitation is an attractive force when the (SEC) holds and the
geodesics get focused as a result of this attraction.\\
 For a congruence of null geodesics be hypersurface orthogonal and (NEC)
  hold, too we have $\frac{d\theta}{d\tau}\leqslant
0$ where $\theta=k^{\alpha}_{;\alpha}$ and $k_{\alpha}$ is the
tangent vector field. It is important to realize that the
Raychaudhuri equation is purely geometric and independent of the
gravity theory under consideration. The connection with the gravity
theory comes from the fact that, in order to relate the expansion
variation with the energy-momentum tensor, one needs the field
equations to obtain the Ricci tensor. Thus, through the combination
of the field equations and the Raychaudhuri equation, one can set
physical conditions for the energy-momentum tensor. The requirement
that gravity is attractive imposes constraints on the
energy-momentum tensors and establishes which ones are compatible.
Of course, this requirement may not hold at all instances. Indeed, a
repulsive interaction is what is needed to avoid singularities as
well as to achieve inflationary conditions, and to account the
observed accelerated expansion of the universe.
\subsection{b : Straightforward calculations}
Now, we consider a congruence of radial, marginally bound, time like
geodesics of the  LB metric \cite{1}:
\begin{eqnarray}
ds^{2}=dr^2+w(r)(-dt^2+d\varphi^2+dz^2)
\end{eqnarray}
Where,
\begin{eqnarray}\nonumber
w(r)=e^{-2\sqrt{\frac{\Lambda}{3}}r}(\xi^2e^{2\sqrt{3\Lambda}r}+1)^{2/3}
\end{eqnarray}
 For radial geodesics,the components of 4-vector field
$u^{\alpha}$, $u^{\varphi}=u^{z}=0$, and the geodesics are
marginally bound if
$-u_{\alpha}\xi^{\alpha}_{(t)}=-u_{t}=\tilde{E}$. This means that
the conserved energy is precisely equal to the rest-mass energy, and
this gives us the equation $u^t=\frac{\tilde{E}}{w(r)}$.\footnote{
Notice that  only in Schwarzschild spacetime (which has the property
$g_{tt}g_{rr}=-1$ )the choose of $\tilde{E}=1$ is a good
  coordinate's  representation and in general ,specially in the form
  of a cylindrically symmetric LB metric, we must keep
  $\tilde{E}\neq1$ to avoiding of  occurrence an unphysical radial
  velocity}
From LB metric  \cite{1}  we know that $w(r)>1$.Indeed since the
radial like component of 4-vector velocity must be a real function
one can deduced that
\begin{eqnarray}\nonumber
 |\tilde{E}|\geq 1
\end{eqnarray}
  From the normalization condition
\begin{eqnarray}\nonumber
g_{\alpha\beta}u^{\alpha}u^{\beta}=-1
\end{eqnarray}
we have:
\begin{eqnarray}\nonumber
u^{r}=\pm\sqrt{\frac{\tilde{E}^2}{w(r)}-1}
\end{eqnarray}
The upper sign applies to outgoing geodesics, and the lower sign
applies to ingoing geodesics.The 4-velocity is given by:
\begin{eqnarray}
u^{\alpha}=(\frac{\tilde{E}}{w(r)},\pm\sqrt{\frac{\tilde{E}^2}{w(r)}-1},0,0)
\end{eqnarray}
And , using (16) we can write:
\begin{eqnarray}\nonumber
u^{\alpha}\partial_{\alpha}=\frac{1}{w(r)}\partial_{t}\pm(\frac{\tilde{E}^2}{w(r)}-1)^{1/2}\partial_{r}\\\nonumber
u_{\alpha}dx^{\alpha}=-\tilde{E}dt\pm(\frac{\tilde{E}^2}{w(r)}-1)^{1/2}
dr
\end{eqnarray}
It follows that $u_{\alpha}$ is equal to a gradient of a scalar
function $\Phi$ where:
\begin{eqnarray}\nonumber
u_{\alpha}=-\Phi_{,\alpha}
\end{eqnarray}
and:
\begin{eqnarray}
\Phi=-t\pm\int(\frac{\tilde{E}^2}{w(r)}-1)^{1/2} dr
\end{eqnarray}
The integral could be written in terms of hypergeometric functions
which we don't write it here .This expression means that the
congruence is everywhere orthogonal to the spacelike hypersurfaces
$\Phi=\emph{constant}$. The expansion is calculated as:
\begin{eqnarray}
\theta=u^{\alpha}_{;\alpha}=\pm
w(r)^{-3/2}\frac{d}{dr}(w(r)\sqrt{\tilde{E}^2-w(r)})
\end{eqnarray}
Not surprisingly, the congruence is \emph{diverging}
\emph{$(\theta>0)$} if the geodesics are\emph{ outgoing}, and
\emph{converging} \emph{$(\theta<0)$} if the geodesics are\emph{
ingoing}.The rate of change of the expansion is calculated as:
\begin{eqnarray}\nonumber
\frac{d\theta}{d\tau}=(\frac{d\theta}{dr}).\frac{dr}{d\tau}=\acute{\theta}u^r
\end{eqnarray}
and the result is:
\begin{eqnarray}
\frac{d\theta}{d\tau}=(\frac{\tilde{E}^2}{w(r)}-1)^{\frac{1}{2}}\frac{d}{dr}[w(r)^{-3/2}\frac{d}{dr}(w(r)\sqrt{\tilde{E}^2-w(r)})]
\end{eqnarray}
Substituting $w(r)$ and performing differentiation , in terms of a
new variables
\begin{eqnarray}\nonumber
x\equiv\frac{w(r)}{\tilde{E}^2}\in(\sqrt[3]{4b^2},1),|b|<\frac{1}{2}\\\nonumber
b\equiv|\frac{\xi}{\tilde{E}}|
  \\\nonumber y\equiv\sqrt{x^6-4b^2x^3}\\\nonumber
a=\sqrt{\frac{3}{\Lambda}}
\end{eqnarray}
We have:
\begin{eqnarray}
\frac{d\theta}{d\tau}=\frac{\Lambda}{2}\frac{\Phi_{b}(x,y)}{x(1-x)}
\end{eqnarray}
Where,
\begin{eqnarray}
\Phi_{b}(x,y)=\frac{(27x^2-45x+20)y-36b^2x^2-46x^4+27x^5+64b^2x+20x^3-32b^2}{3(x^3+y)}
\end{eqnarray}
This polynomial has at most two real roots, as can be seen from the
fact that its second derivative is always positive and it vanishes
whenever there is a double root.As mentioned before, in order to
ensure reality, the root point must be at the right of the outer
region $w(r)<1$.For $|b|<1/2,0<x<1$ this function always poss
negative values.For $\frac{d\theta}{d\tau}=0$ this entails to look
for an stationary point of the equation $\Phi_{b}(x,y)=0$. This
equation can not  be solved.But for suitable values of $b$ this
polynomial in $x$ has a second derivative that is everywhere
positive regardless of the value of $b$, implying that it may have
at most two real roots that can be identified with two \emph{event
horizons}.If one take $b=0$, when the minus sign is chosen,the
metric (15) corresponding to the $LC\Lambda$ case already analyzed
in the previous work[1].These choose will coincide when the roots
become $w(r)=0.377,1.178$ and  a single double root $w(r)=0$(which
is not accessible), by  $w(r)=1.178$ we have
\begin{eqnarray}\nonumber
r=0.0273 a
\end{eqnarray}

 For a congruence of null geodesics which must be hypersurface orthogonal
(in the manner that is discussed in part(a) ) and (NEC) hold, we
must have  $\frac{d\theta}{d\lambda}\leq 0$ where
$\theta=k_{;\alpha}^{\alpha}$ where $k_{\alpha}$ is the tangent
vector field.\\
For $d\varphi=dz=0$ the (LB) line element reduces to:

\begin{eqnarray}\nonumber
ds^2=-w(r)(dt-\frac{dr}{\sqrt{w(r)}})(dt+\frac{dr}{\sqrt{w(r)}})
\end{eqnarray}
The displacements will be \emph{null } if $ds^2=0$. We define two
null coordinates ,
\begin{eqnarray}\nonumber
u=t-r^{*}\\\nonumber v=t+r^{*}
\end{eqnarray}
which as usual we introduced a \emph{tortoise
coordinate}\footnote{$F(a,b,c;z)=\sum_{k=0}^{\infty}\frac{(a)_{k}(b)_{k}}{(c)_{k}
\Gamma(k+1)}z^{k},(a)_{k}=\frac{\Gamma(a+k)}{\Gamma(a)},\Gamma(s)=\int_{0}^{\infty}e^{-t}t^{s-1}dt$}
\begin{eqnarray}\nonumber
r^{*}=\int\frac{dr}{\sqrt{w(r)}}=ae^{\frac{r}{a}}F(\frac{1}{6},\frac{1}{3},\frac{7}{6},-\xi^2e^{\frac{6r}{a}})
\end{eqnarray}
Easily we find that on \emph{out-going} null geodesics
\emph{$u=constant$ } and similarly \emph{$v=constant$ } on
\emph{in-going} ones.The following vectors are  \emph{null},
\begin{eqnarray}\nonumber
k_{\alpha}^{\emph{out}}=-\partial_{\alpha}u\\\nonumber
k_{\alpha}^{\emph{in}}=-\partial_{\alpha}v
\end{eqnarray}
They both satisfies the geodesic equation with $+r$ as an affine
parameter for $k_{\alpha}^{\emph{out}}$ and $-r$  for
$k_{\alpha}^{\emph{in}}$. The congruences are clearly hypersurface
orthogonal.Expansion(s) are calculated:
\begin{eqnarray}
\frac{d\theta}{d\lambda}=\frac{1}{w(r)}\sqrt{\tilde{E}^2-w(r)}[w''(r)-\frac{3}{2}\frac{w'(r)^2}{w(r)}]
\end{eqnarray}
This function never vanishes and remains always negative.We can
construct a similar function as(15) for it and determining the sign
of it.Another simple method is drawing a graph for
$\frac{d\theta}{d\lambda}$.Appling any of this two methods prove
that this function is negative every where.
\section{summary}
In this short report we found the unique exact solution for field
equations containing a massless scalar field and a cosmological
constant term . We checked the stability ,also we showed that  there
is some restriction on the energy conditions.This exact solution
satisfy energy conditions and the validity of focusing theorem
proved directly.

\section{Acknowledgement}
H.R.Rezazadeh thanks the editor of IJTP , Prof. Heinrich Saller and
the anonymous referees made excellent observations and suggestions
which resulted in substantial improvements of the presentation and
the results. The author would like to thank the Islamic Azad
University, Karaj  branch for kind hospitality and support. This
work is supported by Islamic Azad University (Karaj branch). Address
and affiliation at the time this paper was written: Department of
Mathematics, Islamic Azad University, Karaj branch, 31485-313, Iran

\end{document}